\documentclass[12pt]{article}

\usepackage{amsmath,amssymb,amsthm}
\usepackage[margin=2.5cm]{geometry}
\usepackage{hyperref}
\usepackage{xcolor}
\usepackage{graphicx}

\newcounter{Theorems}
\newcounter{Definitions}
\newcounter{Conjectures}
\begin{document}
\begin{titlepage}
\begin{flushright}

\end{flushright}

\begin{center}
{\Large\bf $ $ \\ $ $ \\
Derived brackets in bosonic string sigma-model
}\\
\bigskip\bigskip\bigskip
{\large Vinícius Bernardes, Andrei Mikhailov,  Eggon Viana}
\\
\bigskip\bigskip
{\it Instituto de Fisica Teorica, Universidade Estadual Paulista\\
R. Dr. Bento Teobaldo Ferraz 271, 
Bloco II -- Barra Funda\\
CEP:01140-070 -- Sao Paulo, Brasil\\
}

\vskip 1cm
\end{center}

\begin{abstract}
We study the worldsheet theory of bosonic string from the point of view of the BV formalism.
We explicitly describe the derived Poisson structure which arizes when we expand the Master
Action near a Lagrangian submanifold.
The BV formalism allows us to clarify the mechanism of holomorphic factorization of string
amplitudes.
\end{abstract}

\vfill
{\renewcommand{\arraystretch}{0.8}%
}

\end{titlepage}

\tableofcontents

\section{Introduction}\label{Nonlinearity}

In BV formalism we  study all gauge fixings simultaneously. The Master Action $S_{\rm BV}$
would probably better be called ``Universal Action'' because it covers all possible
gauge fixings. Gauge fixings correspond to choices of a Lagrangian submanifold $L$ in the BV space.
Suppose that we have choosen the Darboux coordinates $\phi^a, \phi_a^{\star}$ so that
$L$ is given by $\phi_a^{\star}=0$. In the vicinity of $L$ we can expand $S_{\rm BV}$
as follows:
\begin{equation}
   S_{\rm BV} = S_0(\phi) + S_1^a(\phi)\phi_a^{\star} + S_2^{ab}(\phi)\phi_a^{\star}\phi_b^{\star}
   + \ldots
   \end{equation}
The coefficients $S_0,S_1,\ldots$ can be equivalently described
as $\infty$-Poisson structures \cite{Voronov:2005}, a generalization of Poisson brackets. 
For functions $F_1(\phi),\ldots F_n(\phi)$ (which depend on $\phi$, not $\phi^{\star}$;
                                                    they are functions on $L$)
their $n$-bracket is defined as:
\begin{equation}
   \{F_1,\ldots,F_n\} = \{\ldots\{\{S_{\rm BV},F_1\}_{\rm BV},F_2\}_{\rm BV},\ldots F_n\}_{\rm BV}|_{\phi^{\star}  = 0}
   \end{equation}
The derived brackets describe the dependence of $S_{\rm BV}|_L$ on $L$.
Suppose that we deformed $L$:
\begin{equation}
     \phi^{\star}_a = 0 \stackrel{\rm deform}{\longrightarrow} \phi^{\star}_a = {\partial \Psi(\phi)\over\partial \phi^a}
     \label{DeformLAG}\end{equation}
Then:
\begin{equation}
   S_{\rm BV}|L = S_0 + Q\Psi + {1\over 2}\{\Psi,\Psi\} + {1\over 6}\{\Psi,\Psi,\Psi\} + \ldots
   \end{equation}
The Master Equation implies for these brackets some 
generalized Jacobi identities. It is interesting to observe
how they are satisfied in special cases. Especially in the case of bosonic string,
where the deformations (\ref{DeformLAG}) play crucial role.
Indeed, string perturbation theory is defined as integral over families
of such  deformations \cite{Schwarz:2000ct}.

When $S_0 = 0$ the $\infty$-Poisson structure becomes  \emph{strict}.
However, it is possible to have a strict Poisson structure with $S_0\neq 0$.
It is sufficient that BV brackets of $S_n$ with $S_0$ vanish.
When a homotopy Poisson structure is not strict, this means that the homotopy Jacobi identities
are satisfied only modulo the derivatives of $S_0$ (``on-shell'').
Generally speaking, there is no sense in which the brackets preserve the extremal set of $S_0$,
and therefore no obvious way to turn a non-strict structure into a strict one.

In this paper we will consider the derived Poisson brackets which arize in the worldsheet
theory of bosonic string. This theory is of the type ``BV-BRST'';
the BRST structure corresponds to worldsheet diffeomorphisms, which are gauge symmetries.
(In some formulations, there is also Weyl gauge symmetry.)
When Lagrangian submanifold is choosen in the usual way, the BRST operator is only
nilpotent on-shell, and there is an infinite tower $S_0, S_1^{a}, S_2^{ab}, \ldots$, generating
a homotopy Poisson structure. This ``usual'' choice of a Lagrangian submanifold is a particular
case of the general construction of BRST formalism, where we fix a constraint and choose
a Lagrangian submanifold  as a conormal bundle of the constraint surface.
In this general context, we describe a way to make the homotopy Poisson structure strict,
by imposing certain conditions  on the Faddeev-Popov ghosts.
In the case of bosonic string, it turns out that the differential actually commutes with
the derived brackets, and the derived brackets are in some sense constant.
We observe that this holds in general, in the BV-BRST formalism, under the condition
of vanishing of some cohomological obstacles, see
Section \ref{HigherPoissonBrackets} and
Section \ref{BosonicString}.
 
In the case of the bosonic string, the most important example of the deformations
of Lagrangian submanifold are those which correspond to variations of the worldsheet
complex structure. When the target space is flat, we show that Master Equations
imply effective linearization of the deformation of the complex structure, see
Section \ref{DerivedBracketsAndContactTerms}.
As was explained in \cite{DHoker:2015qrj}, this effective linearization
plays crucial role in the holomorphic factorization of string amplitudes
\cite{Belavin:1986cy},\cite{Verlinde:1986kw}.

\section{Brief review of BV-BRST formalism}\label{ReviewBRST}

         Bosonic string worldsheet theory belongs to a class of gauge systems which can
         be called ``BV-BRST''.
         It is a particular case of BV formalism, when fields/antifields can be choosen so that
         the expansion of the Master Action in powers of antifields terminates at the linear term:
         \begin{equation}
            S_{\rm BV}(\Phi, \Phi^{\star}) = S_{\rm cl}(\Phi) + Q^A(\Phi)\Phi_A^{\star}
            \end{equation}
         Moreover, the odd nilpotent vector field $Q^A$ should be of some special form,
         coming from some gauge symmetry of $S_{\rm cl}(\Phi)$, as we will now review.

\subsection{Master Action in BRST case}\label{sec:MasterActionInBRST}

Let $H$ be the gauge group with
gauge Lie algebra $\bf h$, acting on the space of fields $\phi$.
We enlarge the space of fields, by adding ``ghosts'' $c$ parametrizing $\Pi {\bf h}$
(the Lie algebra of $\bf h$ with flipped statistics).
The ``total'' space of fields is now $\Pi  {\bf h} \times X$, where $\Pi \bf h$ is parametrized by $c$
and $X$ by $\phi$. The Master Action has the form:
\begin{equation}
   S_{\rm BV}(\phi,\phi^{\star}) = S_{\rm cl}(\phi) + c^A v_A^i(\phi)\phi^{\star}_i +
   {1\over 2} c^A c^Bf_{AB}^C c^{\star}_C
   \end{equation}
In the case of bosonic strings, gauge symmetries are diffeomorphisms, and $\bf h$
is the algebra of vector fields on the worldsheet.

For quantization we need to restrict $S_{\rm BV}$ to a Lagrangian submanifold,
and then take the path integral over the fields which parametrize the Lagrangian submanifold.

\subsection{Construction of Lagrangian submanifolds by constraining fields}\label{sec:LAGByConstraints}

How do we choose a Lagrangian submanifold? A very naive choice is $\phi^{\star}=0$,
then the restriction would be just $S_{\rm cl}$ --- degenerate, can not quantize.

One solution is to use the ``conormal bundle'' construction, which we will now describe. 
Let $\cal F$ denote the space
of fields $\phi$. Consider some subspace ${\cal C}\subset {\cal F}$ defined by some equations
(``constraints''). 
We choose:
\begin{equation}
L \;=\; \Pi(T{\cal C})^{\perp}\times \mbox{[$c$-ghosts]}\;=\; \Pi(T{\cal C})^{\perp}\times \Pi {\bf h}
\label{rotated-lag}\end{equation}
where $T{\cal C}^{\perp}\subset T^*{\cal F}$ consists of those linear functionals which
vanish on the tangent space to $\cal C$.
This is a Lagrangian submanifold. 
We want $\cal C$ to be ``sufficiently transverse'' to the the gauge orbits,
for the resulting action to be sufficiently non-degenerate.

\section{Derived Poisson brackets in BRST/BV formalism}\label{HigherPoissonBrackets}

         Now we will consider the expansion of the Master Action in the
         vicinity of a Lagrangian submanifold corresponding to a constraint.
         Such an expansion always defines an $\infty$-Poisson structure,
         but in the BRST case we show that it can be made \emph{strict}
         by imposing certain conditions on ghosts.
         Moreover, under the condition of vanishing of some cohomological obstacle,
         the higher brackets can be made essentially constant by a choice of coordinates.

\subsection{BV phase space in the vicinity of a constraint surface}\label{sec:VicinityOfConstraintSurface}

Suppose that we can choose coordinates $x^m,m^I$ on $\cal F$ in the vicinity of $\cal C$
so that the equation of $\cal C$ is:
\begin{equation}
   m^I = 0
   \end{equation}
This actually defines a family of surfaces ${\cal C}_{(m_0)}$ given by equations:
\begin{equation}
   m^I = m_0^I
   \end{equation}
For each ${\cal C}_{(m_0)}$,
the differentials $dm_I$ form the basis of the fiber of the conormal
bundle of ${\cal C}_{(m_0)}$.
Any element of the fiber can be written as:
\begin{equation}
   b_I dm^I
   \end{equation}
The odd symplectic form is:
\begin{equation}
   \omega_{\rm BV} \;=\; db_I\, dm^I \,+\, dx_{\mu}^{\star}\, dx^{\mu}
   \end{equation}
The BV action is:
\begin{align} S_{\rm BV} \;=\;
 &S_{\rm cl}(x,m) + c^Av_A^I(x,m) b_I \;+\nonumber{} \\  
 &+ \; c^Av^{\mu}_A(x,m) x^{\star}_{\mu} + {1\over 2}[c,c]c^{\star}\label{BVBRST} \end{align}
Only the first line contributes to the restriction on $L = \Pi T{\cal C} \times \Pi {\bf h}$:
\begin{equation}
   S_0 = \left. S_{BV} \right|_L = S_{\rm cl}(x,0) + c^Av_A^I(x,0) b_I
   \label{GeneralS0}\end{equation}

\subsection{Restriction on ghosts}\label{sec:RestrictionOnGhosts}

We allow for the possibility that constraining fields to ${\cal C}\subset {\cal F}$
does not fix the gauge symmetry completely, but actually fixes it to some subgroup
$H_{\cal C}\subset H$. 

For example, in the case of bosonic string,
we break diffeomorphisms down to conformal transformations.

We can formally restrict the group  of gauge transformations to $H_{\cal C}\subset H$, in the following way.
Consider the subspace ${\bf h}_{\cal C}^{\perp}\subset {\bf h}^*$ consisting of those linear functions which
vanish on ${\bf h}_{\cal C}$. Consider the subspace of the space of fields defined by
the following constraint:
\begin{equation}
   f_A c^A = 0 \quad\forall\quad f\in {{\bf h}_{\cal C}^{\perp}\subset {\bf h}^*}
   \label{PartialOnShell}\end{equation}
Notice that $Q$ preserves this constraint, because ${\bf h}_{\cal C}$ is a Lie subalgebra.

At the same time, we consider  $c^{\star}$ mod $\Pi{\bf h}_{\cal C}^{\perp}$.

This restriction on ghosts might be thought of, in some sense, as a partial on-shell condition
(only those EqM which follow from variation over $b$).
We denote the reduced field space ${\cal F}_{\rm res}$:
\begin{equation}
   {\cal F}_{\rm res} \subset {\cal F}
   \end{equation}
The subindex ``res'' stands for ``residual'', because we are essentially restricting ghosts
to the algebra of residual
(\textit{i.e.} those remaining after we impose the constraint ${\cal C}\subset {\cal F}$)
gauge transformations.

This restriction on ghosts may be interpreted as a ``partial on-shell condition'',
but only vaguely.
Although we have chosen a Lagrangian submanifold $L$, we are actually working
in the BV space (in some vicinity of $L$). There is no good notion of equations of motion,
or ``on-shell'', in the BV space. What we really use is the fact that ghosts are
separate from other fields, as we are in the BRST context. Imposing Eq. (\ref{PartialOnShell})
is geometrically natural in this context. It is a ``BV Hamiltonian reduction'' of the BV phase space
on the constraint given by Eq. (\ref{PartialOnShell}), in other words $\Pi T^* {\cal F}_{\rm res}$.
It happens to coincide with the equations of motion
from the variation of $b$, in restriction to $L$.

\subsection{$\infty$-Poisson structure}\label{sec:InfinityPoissonStructure}

\subsubsection{Expansion of $S_{\rm BV}$ in the vicinity of $L$}\label{para:ExpansionOfSBV}

We can expand $S_{\rm BV}$ in the vicinity of $L$:
\begin{equation}
   S_{\rm BV} = S_0 + S_1 + S_2 + \ldots
   \end{equation}
where the subindex $0,1,2,\ldots$ counts the degree in $m$ (antifield to $b$)
plus the degree in $x^{\star}$ plus the degree in $c^{\star}$. 
Or, once we impose Eq. (\ref{PartialOnShell}):
\begin{equation}
   S_{\rm BV, res} = S_{0,\rm res} + S_{1,\rm res} + S_{2,\rm res} + \ldots
   \end{equation}
Once we impose Eq. (\ref{PartialOnShell}), the $S_0$ becomes $b$-independent
(\textit{cp.} Eq. (\ref{GeneralS0})):
\begin{equation}
   S_{0,{\rm res}} = S_{\rm cl}(x,0)
   \end{equation}
We have, because of the Master Equation:
\begin{equation}
   \{S_{0,{\rm res}}\,,\,S_{1,{\rm res}}\}_{\rm BV}  = 0
   \end{equation}
Moreover, since $x^{\star}$ only enters in $S_1$,
and ${\delta\over\delta b}S_{0,{\rm res}}=0$, we have:
\begin{equation}
   \{S_{0,{\rm res}}\,,\,S_{n,{\rm res}}\}_{\rm BV}  = 0
   \label{SnCommuteWithS0}\end{equation}
Let us denote:
\begin{equation}
   Q_{\rm res} = \{S_{1,\rm res},\_\}_{\rm BV}
   \end{equation}
Then Eq. (\ref{SnCommuteWithS0}) implies:
\begin{equation}
   Q_{\rm res} S_{n,\rm res} + {1\over 2} \sum_{k=2}^{n-1} \{S_{k,\rm res}\,,\,S_{n+1-k,\rm res}\}_{\rm BV} = 0
   \end{equation}
In particular $Q_{\rm res}$ is nilpotent:
\begin{equation}
   Q_{\rm res}^2 = 0
   \end{equation}
\emph{Summary} We constructed a tower of brackets $\pi_n$ on a space parameterized
by $x,c,b$ which form an $\infty$-Poisson structure.
The brakets $\pi_{\geq 2}$  only involve derivatives in the $b$-direction,
while $Q=\pi_1$ contains also derivatives w.r.to $x$ and $c$.

\subsection{BRST symmetry}\label{sec:BRST}

For any Lagrangian submanifold $L$, the restriction of $S_{\rm BV}$ to $L$ has
a fermionic symmetry $Q_{\rm BRST}$ which is defined as follows. Let us introduce
Darboux coordinates so that the equation for $L$ is: $\phi^{\star}_a = 0$.
Let us expand $S_{\rm BV}$ in powers of $\phi^{\star}$:
\begin{equation}
   S_{\rm BV} = S_0(\phi) + Q^a_{\rm BRST}(\phi) \phi^{\star}_a + \ldots
   \end{equation}
Master Equation implies that $Q^a(\phi){\partial\over\partial\phi^a}$ is a symmetry of $S_0$.
It is nilpotent, generally speaking, only on-shell.

\subsection{A condition for simplification of $\infty$-Poisson structure}\label{sec:ConditionForSimplification}

Consider the Master Action:
\begin{equation}
   S_{\rm BV} \;=\; S_{\rm cl}(x,m) + c^Av_A^I(x,m) b_I +
   c^Av^{\mu}_A(x,m) x^{\star}_{\mu} + {1\over 2}[c,c]c^{\star}
   \end{equation}
in the vicinity of the Lagrangian submanifold $L$ given by the equation:
\begin{equation}
   m = x^{\star} = c^{\star} = 0
   \end{equation}
We restrict $c$ to a subalgebra of ${\bf h}_{\cal C}\subset {\bf h}$ preserving $L$. Explicitly, the condition on $c$ is:
\begin{equation}
   c^Av_A^I(x,0) = 0
   \end{equation}
Suppose that the following equation is also true, for some $x_0$:
\begin{equation}
   c^Av_A^{\mu}(x_0,0) = 0
   \end{equation}
In other words, suppose that $x= x_0$ is a fixed point of ${\bf h}_{\cal C}$.
Then we have a \emph{vector field}:
$
   v\langle c\rangle = c^Av_A^I(x,m) {\partial\over\partial m^I}
   +
   c^Av^{\mu}_A(x,m) {\partial\over\partial x^{\mu}}
   $
vanishing at $x=m=0$ and satisfying the  Maurer-Cartan equation:
\begin{equation}
   [c,c]{\partial\over\partial c} v\langle c\rangle
   +
   \left[
         v\langle c\rangle , v\langle c\rangle 
         \right]
   \end{equation}
It
\href{https://andreimikhailov.com/math/Renormgroup/sec_Linearization.html}{\textbf{\textcolor{blue}{was proven}}}
in \cite{Mikhailov:2019kfm} that under certain conditions we can adjust the coordinates
in such a way that the vector fields $v_A$ become linear in $x$ and $m$.
Potential obstacles are the cohomology groups:
\begin{equation}
   H^1\left({\bf h}_{\cal C}\,,\, \mbox{Vec}_{\geq 2}(x,m)\right)
   \end{equation}
where $\mbox{Vec}_{\geq 2}(x,m)$ are polynomial vector fields on $(x,m)$-space of quadratic and higher order,
on which ${\bf h}_{\cal C}$ act by \emph{linear} vector fields $v_A$.
Suppose that the obstacles vanish, and we  can indeed make $v_A$ linear in $x$ and $m$.
Then it follows that the $b$-dependent part of $\pi_n$ vanishes for $n>1$.
Although higher brackets are nonzero, they are all $Q$-closed.
Moreover, since $\pi_n$ only contain derivatives in the $b$ direction, they are
therefore ``essentially constant''.

This is what happens in the case of bosonic string, which we will describe in
Section \ref{BosonicString} --- see Eq. (\ref{ConformalTransformationOfM}).

\subsection{Changes of coordinates}\label{sec:ChangesOfCoordinates}

Let us consider the change of coordinates $m$ (which parametrize the choice of constraints):
\begin{equation}
   m^I = m^I(\mu)
   \end{equation}
Then:
\begin{align} b_I\, dm^I \;=\;
 &b_I {\partial m^I\over\partial \mu^a} d\mu^a = \beta_a d\mu^a\nonumber{} \\ \mbox{where \hspace{1.00000ex}}
 &\beta_a = b_I {\partial m^I\over\partial \mu^a}\nonumber{} \end{align}
This defines the change of Darboux coordinates from $b_I,m^I$ to $\beta_a,\mu^a$:
\begin{equation}
   db_I\, dm^I \;=\; d\beta_a\, d\mu^a
   \end{equation}

\section{Bosonic string}\label{BosonicString}

         We will now apply the general BV/BRST formalism to the case of bosonic string
         worldsheet theory,
         following \cite{Craps:2005wk},\cite{Mikhailov:2016myt},\cite{Mikhailov:2016rkp}.
         As a constraint, we will choose fixing the complex structure $I$ of the string
         worldsheet to some particular ``c-number'' complex structure $I^{(0)}$.
         This defines some Lagrangian submanifold as described in
         Section \ref{sec:LAGByConstraints}.
         Then in
         Section \ref{sec:ComplexStructureConstraint}
         we find some BV Darboux coordinates, adjusted to this Lagrangian submanifold,
         and study the expansion of the Master Action in powers of antifields.
         Then in
         Section \ref{sec:MuMuBar} 
         we change the Darboux coordinates so that one of the antifields becomes the
         Beltrami differential.
         We find that the derived brackets have a very particular structure:
         all higher brackets are essentially constant.

\subsection{Polyakov gauge in BV formalism}\label{sec:PolyakovGauge}

The bosonic string is described by the Polyakov action
\begin{align}  
 &S_{cl} = \frac{1}{4 \pi \alpha'} \int_{\Sigma} dx \wedge \star dx = \frac{1}{4 \pi \alpha'} \int dx \wedge I \cdot dx,\label{classical_action} \end{align}
where the dynamical fields are $D$ scalar fields $x^0,\ldots, x^{D-1}$ and a complex structure $I$ on the worldsheet. The complex structure is a section of $\mbox{End}(T\Sigma)$ satisfying $I^2 = -1$ (the integrability condition is automatically satisfied in two dimensions).
It defines the Hodge operator $\star$, which takes the 1-form $dx$ to the 1-form $\star dx$.

The action has gauge symmetry under diffeomorphisms. We therefore consider the
diffeomorphism ghosts $c$ (corresponding to all smooth vector fields on the worldsheet), 
the BRST operator $Q$, as the  generating function $\hat{Q}$ of $Q$ on the BV phase space:
\begin{align}  
 &\hat Q = \int ( (\mathcal{L}_c x) x^{\star} + (\mathcal{L}_c I) I^{\star} + \frac{1}{2} (\mathcal{L}_c c) c^{\star} ),\label{classical_brst} \end{align}
where $x^\star$, $I^\star$ and $c^\star$ are the antifields. A BV action can be defined as
\begin{align}  
 &S = S_{cl} + \hat Q\nonumber{} \end{align}
It satisfies the quantum master equation $\{ S, S \} = 0$, where the bracket is the BV bracket.

We will choose the Lagrangian submanifold given by $x^{\star}=0$, $c^{\star}=0$, and $I = I^{(0)}$
for some fixed complex structure $I^{(0)}$. We will construct the BV Darboux coordinates adjusted
to this choice of Lagrangian submanifold, in the following way.
The ``fields'' will be $x$, $c$, and $I^{\star}$. The antifields will be
$x^{\star}$, $c^{\star}$ and something like  $I - I^{(0)}$, see 
Section \ref{sec:ComplexStructureConstraint}.

\subsubsection{Odd cotangent bundle of the space of complex structures}\label{sec:ComplexStructureConstraint}

Consider the space of 2x2 matrices $I$ satisfying $I^2=-1$. Let $I^{\star}$ be
a 2x2 matrix parametrizing the fiber of the odd cotangent bundle to the space of $I$'s.
The odd symplectic form is:
\begin{equation}
   \omega = \mbox{tr}(dI^{\star}\wedge dI)
   \label{OddSymplecticFormI}\end{equation}
with the following gauge symmetry of $I^{\star}$:
\begin{equation}
   \delta_{\eta} I^{\star} = \eta I + I \eta
   \end{equation}
Let us gauge fix $I^{\star}$ to:
\begin{equation}
   I^{\star} = \frac{1}{2} \left( \begin{array}{cc} 0 & i \bar{b} \cr - ib & 0 \end{array} \right)
   \label{parameterization_bfield}\end{equation}
The fields $b$ and $\bar{b}$ are called ``$b$-ghosts''.

A generic $I$ satisfying $I^2 =  -1$ can be parametrized by
$m\in {\bf C}$:
\begin{align}  
 &I =
          \begin{pmatrix}
          I^z_z & I^z_{\bar{z}} \\ I^{\bar{z}}_z & I^{\bar{z}}_{\bar{z}}
          \end{pmatrix}
          =
          \begin{pmatrix}
          i \sqrt{ 1 + m \bar m } & im \\
          -i \bar m & - i \sqrt{ 1 + m \bar m }
          \end{pmatrix}.\label{ParametrizationOfI} \end{align}
Notice that in our notations, the reality conditions are:
\begin{equation}
     \begin{pmatrix} 0 & 1 \\ 1 & 0 \end{pmatrix}
     I
     \begin{pmatrix} 0 & 1 \\ 1 & 0 \end{pmatrix}
     =
     \overline{I}
     \quad , \quad
     \begin{pmatrix} 0 & 1 \\ 1 & 0 \end{pmatrix}
     I^{\star}
     \begin{pmatrix} 0 & 1 \\ 1 & 0 \end{pmatrix}
     =
     \overline{I^{\star}}
     \label{RealityConditionsI}\end{equation}
Eq. (\ref{OddSymplecticFormI}) becomes:
\begin{equation}
   \omega = \frac{1}{2} d m\,d b + \frac{1}{2} d \bar{m}\, d \bar{b}
   \label{OddSymplecticFormIFixed}\end{equation}
Therefore, $m$ and $b$ are Darboux coordinates. 
Remember that $I$ and $I^{\star}$ are actually fields on the string worldsheet. Including
also $x$, $c$ and their antifields, we get:
\begin{align} \omega =
 &~\int \left( \delta x \delta x^\star + \delta c \delta c^\star + \delta \bar c \delta \bar c^\star + \frac{1}{2} \delta m \delta b + \frac{1}{2} \delta \bar m \delta \bar b \right).\nonumber{} \end{align}
The corresponding odd Poisson bracket is:
\begin{align} \{ b (z) , m (z') \} =
 &~ 2 \delta^2(z - z'), \hspace{1cm}
             \{ b (z), \bar m (z') \} = 0,\nonumber{} \\ \{ \bar b (z), m (z') \} =
 &~ 0,\hspace{1cm}
             \{ \bar b (z), \bar m (z') \} = 2 \delta^2 (z - z').\label{darboux_bracket} \end{align}

\subsubsection{Expansion of the action}\label{sec:ExpansionOfAction}

Let us consider the expansion of the Master Action:
\begin{equation}
   S_{\rm BV} = S_{\rm cl} + \widehat{Q}
   \end{equation}
around the Lagrangian sumbamifold $x^{\star} = c^{\star} = m = 0$.
The $S_{\rm cl}$ given by Eq. (\ref{classical_action})) is written in terms of the $m, \bar m$ fields as:
\begin{align} S_{cl} =
 &\int d^2 z \left( \sqrt{ 1 + m \bar m }
                                  \partial x \bar \partial x + \frac{1}{2} m ( \partial x )^2 + \frac{1}{2} \bar m ( \bar \partial x )^2 \right),\label{polyakov_darboux} \end{align}
and the BRST operator can be written in terms of $m, \bar m$ and $b, \bar b$ as

\begin{align} \hat Q =
 &~\int d^2 z \bigg( \mathcal{L}_c x x^\star + \frac{1}{2} \mathcal{L}_c c c^\star + \sqrt{1 + m \bar m} ( (\bar \partial c) b +  (\partial \bar c) \bar b ) \;+\nonumber{} \\  
 &\phantom{~\int d^2 z }
              + \frac{1}{2}
              \left(
                    (c \stackrel{\leftrightarrow}\partial m) b +
                    \bar{\partial}(\bar c  m) b  +
                    (\bar{c} \stackrel{\leftrightarrow}{\bar\partial} \bar m)\bar{b} +
                    \partial (c\bar{m}) \bar{b}
                    \right)
              \bigg)\label{QHat} \end{align}
The derivation of this formula uses the following expressions for the components of the
Lie derivative of the complex structure (see Eq. (\ref{ParametrizationOfI})):
\begin{align} (\mathcal{L}_cI)^z_{\ z} \;=\;
 &ic\cdot\partial(\sqrt{1+m\bar{m}}) + i(\bar{\partial}c)\bar{m} + (\partial \bar{c})im\nonumber{} \\ (\mathcal{L}_cI)^z_{\ \bar{z}} \;=\;
 &ic\cdot\partial (m) + 2i(\bar{\partial}c)\sqrt{1+m\bar{m}} + i(\bar{\partial}\bar{c}-\partial c)m\nonumber{} \\ (\mathcal{L}_cI)^{\bar{z}}_{\ z} \;=\;
 &-i c\cdot\partial(\bar{m}) -2i(\partial\bar{c})\sqrt{1+m\bar{m}} -i (\partial c - \bar{\partial}\bar{c})\bar{m}\nonumber{} \\ (\mathcal{L}_cI)^{\bar{z}}_{\ \bar{z}} \;=\;
 &-ic\cdot\partial(\sqrt{1+m\bar{m}}) - i(\partial\bar{c})m - i (\bar{\partial}c)\bar{m}\nonumber{} \end{align}
\emph{If} we restrict the ghost field to be a conformal Killing vector, \textit{i.e.}
$\bar{\partial}c = \partial\bar{c} = 0$, then $m$ transforms 
as a tensor $m^z_{\bar{z}}$:
\begin{equation}
   \{\widehat{Q}, m\}_{\rm BV} = c\stackrel\leftrightarrow\partial m + \bar{\partial}(\bar{c}\bar{m})
   \label{ConformalTransformationOfM}\end{equation}
--- a linear transformation
\footnote{
This is because we do not impose any equations on $m$;
it can be any function of $z,\bar{z}$.
Had we imposed, say, some wave equations, obstacles to linearization could have
appeared as in \cite{Mikhailov:2019kfm}.
}

Now we are ready to play our ``field-antifield flip''. We call $(x,c,b)$ fields
and $(x^{\star},c^{\star},m)$ antifields.
(We call $b, \bar b$ the fields of the Polyakov gauge, and let $m, \bar m$ be the antifields.)
Then we consider the expansion of
$S_{\rm cl} + \widehat{Q}$ around the Lagrangian submanifold where $x^{\star}$, $c^{\star}$, $m$ are all zero. We have:
\begin{align} S =
 &~ S_0 + S_1 + S_2 +\cdots,\nonumber{} \end{align}
where the term $S_k$ is of $k$th power on $x^\star$, $c^\star$ and $m, \bar m$.
The antifields $x^{\star}$ and $c^{\star}$ only enter linearly, but the dependence
on $m$ and $\bar{m}$ is nonlinear.

The term $S_0$ is the string worldsheet action in the Polyakov gauge:
\begin{align} S_0 =
 &~\int d^2 z \left( \partial x \bar \partial x - b \bar \partial c - \bar b \partial \bar c \right)\label{ActionInPolyakovGauge} \end{align}
The first order term $S_1$ is:
\begin{align} S_1 =
 &~\int d^2 z\bigg( \mathcal{L}_c x x^\star + \frac{1}{2} \mathcal{L}_c c c^\star +
                                   \frac{1}{2} m (\partial x)^2 + \frac{1}{2} \bar m (\bar \partial x)^2\nonumber{} \\  
 &~\hspace{1.5cm}
              - \frac{1}{2} b ( c \partial  + \bar c \bar \partial - (\partial c) + (\bar \partial \bar c) ) m
              - \frac{1}{2} \bar b ( c \partial + \bar c \bar \partial - (\bar \partial \bar c) + (\partial c) ) \bar m \bigg).\label{S1} \end{align}
It defines the BRST operator in Polyakov gauge, which generates the BRST symmetry of $S_0$:
\begin{align} Q_{\rm BRST} x \;=\;
 &{\cal L}_c x\nonumber{} \\ Q_{\rm BRST} c \;=\;
 &{1\over 2}{\cal  L}_c c\nonumber{} \\ Q_{\rm BRST} b \;=\;
 &- (\partial x)^2 + 2(\partial c)b + (c\partial  + \bar{c}\bar{\partial}) b = - (\partial x)^2 + {\cal L}_c b\label{QBRSTb} \\ Q_{\rm BRST} \bar{b} \;=\;
 &- (\bar{\partial} x)^2 + 2(\bar{\partial} \bar{c})\bar{b} + (c\partial + \bar{c}\bar{\partial})\bar{b} = - (\bar{\partial} x)^2 + \overline{{\cal L}_c b}\label{QBRSTbbar} \end{align}
Notice that $Q_{\rm BRST} b$ contains the term $\bar{c}\bar{\partial}b$,
and $Q_{\rm BRST}\bar{b}$ contains $c\partial \bar{b}$, which both are zero on-shell.
But, if we omitted them, $Q_{\rm BRST}$ would not be a symmetry of the action.
In this sense, it is not completely accurate to say that BRST variation of $b$
is the energy-momentum tensor.

\subsubsection{On-shell restriction of ghosts}\label{sec:PartialOnShellString}

The only equations of motion required for the nilpotence of this $Q_{\rm BRST}$
are $\partial\bar{c} =0$ and $\bar{\partial}c = 0$.
(The derivation of the nilpotence uses ${\cal L}_c \partial x = \partial {\cal L}_c x$,
     which is only true when $c$ in a conformal Killing vector field.)
Furthermore, if we restrict the ghosts to satisfy
$\partial\bar{c} =0$ and $\bar{\partial}c = 0$,
then $m$ and $\bar{m}$ transform as sections
of $T^{1,0}\otimes \Omega^{0,1}$ and $T^{0,1}\otimes \Omega^{1,0}$, respectively,
in the following sense:
\begin{align} \{S_{\rm BV},m\} \;=\;
 &c\stackrel{\leftrightarrow}{\partial}m + \bar{\partial}(\bar{c}m)\label{TransformationOfM} \\ \{S_{\rm BV},\bar{m}\} \;=\;
 &\bar{c}\stackrel{\leftrightarrow}{\bar\partial}\bar{m} + \partial(c\bar{m})\label{TransformationOfMBar} \end{align}
In particular $\bar{m}m$ transforms as a scalar:
\begin{equation}
     \{S_{\rm BV}, |m|^2\} =  (c\partial + \bar{c}\bar{\partial}) |m|^2
     \label{TransformationOfMMbar}\end{equation}
(This is true only when $c$ and $\bar{c}$ are restricted to 
      $\partial\bar{c} =0$ and $\bar{\partial}c = 0$.)

\subsubsection{Expansion in powers of $m$}\label{sec:ExpansionInPowersOfM}

The quadratic terms are:
\begin{align} S_2 =
 &~\int d^2 z \left( \partial x \bar \partial x - b \bar \partial c - \bar b \partial \bar c \right) {m\bar{m}\over 2}\label{S2} \end{align}
The term of the order $2n$ is, when $n\geq 2$:
\begin{align} S_{2n} =
 &~\int d^2 z \left( \partial x \bar \partial x - b \bar \partial c - \bar b \partial \bar c \right) (-1)^{n+1} {(2n-3)!\over 2^{2n-2} n!(n-2)!} (m\bar{m})^n\label{nth_order} \end{align}

\subsection{Beltrami differential}\label{sec:MuMuBar}

         In order to find the Darboux coordinates adjusted to our choice of Lagrangian submanifold,
         it was useful to parametrize the complex structures by $m,\bar{m}$,
         see Eq. (\ref{ParametrizationOfI}).
         But it is more useful to work, instead of $m$, with Beltrami differentials
         $\mu$, which are holomorphic coordinates on the space of complex structures.
         We will now change the Darboux coordinates from $m,b,\bar{m},\bar{b}$
         to $\mu,\beta,\bar{\mu},\bar{\beta}$, and describe
         the expansion of the Master Action in powers of $\mu,\bar{\mu}$.

\subsubsection{Deformation of Dolbeault operators}\label{sec:DeformationOfDolbeault}

In $z \bar z$ coordinates we can write the complex structure locally as
\begin{align}  
 &I =
    \begin{pmatrix}
        i & 0 \\
        0 & -i
    \end{pmatrix},\nonumber{} \end{align}
so it has eigenvalues $\pm i$.

We define the Dolbeault operators as
\begin{align} \partial :=
 &~ dz\frac{\partial}{ \partial z }, \hspace{1cm}
    \bar \partial := ~ d \bar z \frac{\partial}{ \partial \bar z },\nonumber{} \end{align}
which satisfy $d = \partial + \bar \partial$, where $d$ is the de Rham operator. Their action on the $X$ fields are eigenfunctions of the complex structures, as
\begin{align} I \partial X =
 &~ i\partial X, \hspace{1cm}
    I \bar \partial X = - i \bar \partial X.\nonumber{} \end{align}
Then we can write Polyakov action in terms of $\partial$ and $\bar \partial$, which gives
\begin{align} S =
 &~\frac{ - i }{ 2 \pi \alpha' } \int \partial X\wedge \bar \partial X,\nonumber{} \end{align}
which has the form of the Polyakov action with a flat metric.

To parameterize locally the possible gauge fixing choices we can fix the complex structures to different values. It should be equivalent to fixing the complex structure to be flat in different coordinates.

Let's parameterize the family of complex structures by functions $\mu$ and $\bar \mu$, and call the complex structure $I^{[\mu]}$. For $\mu = \bar \mu = 0$ we get the flat one, $I^{[0]}$. The Dolbeault operators $\partial$ and $\bar \partial$ are eigenvectors of $I^{[0]}$. We can then define deformed Dolbeault operators by
\begin{align} I^{[\mu]} \partial^{[\mu]} =
 &i\partial^{[\mu]},\nonumber{} \\ I^{[\mu]} \bar \partial^{[\mu]} =
 &~ - i\bar \partial^{[\mu]}.\nonumber{} \end{align}
We fix their normalization by asking that
\begin{align}  
 &d = \partial^{[\mu]} + \bar \partial^{[\mu]}\label{derham_dolbeault} \end{align}
continues to hold.

The way we want to parameterize $\partial^{[\mu]}$ and $\bar \partial^{[\mu]}$ is the following. Define the vector fields
\begin{align} v^{[\mu]} =
 &~\frac{ \partial }{ \partial z } + \bar \mu \frac{ \partial }{ \partial \bar z },\nonumber{} \\ \bar v^{[\mu]} =
 &~\frac{ \partial }{ \partial \bar z } + \mu \frac{ \partial }{ \partial z } .\label{SplitOfT} \end{align}
The Dolbeault operators are mixed 2-tensors, so we define them as
\begin{align} \partial^{[\mu]} =
 &~ u^{[\mu]} v^{[\mu]}, \hspace{1cm}
    \bar \partial^{[\mu]} = \bar u^{[\mu]} \bar v^{[\mu]},\nonumber{} \end{align}
for a family of forms $u^{[\mu]}$ and $\bar u^{[\mu]}$. Upon the definition of eq. (\ref{SplitOfT}), the forms are fixed by Eq. (\ref{derham_dolbeault}). They are given by
\begin{align} u^{[\mu]} =
 &~\frac{ dz - \mu d \bar z }{ 1 - \mu \bar \mu  }, \hspace{1cm}
    \bar u^{[\mu]} = \frac{ d \bar z - \bar \mu dz }{ 1 - \mu \bar \mu  }.\nonumber{} \end{align}

The action of the Dolbeault operators on functions is given by:
\begin{align} \partial^{[\mu]}f =
 &~\frac{1}{ 1 - \mu \bar \mu } ( dz - \mu d \bar z ) \left( \frac{ \partial }{ \partial z } + \bar \mu \frac{ \partial }{ \partial \bar z } \right)f,\nonumber{} \\ \bar \partial^{[\mu]}f =
 &~\frac{1}{ 1 - \mu \bar \mu } ( d \bar z - \bar{\mu} d z ) \left( \frac{ \partial }{ \partial \bar z } + \mu \frac{ \partial }{ \partial z } \right)f .\label{dolbeault_mu} \end{align}

\paragraph{Relation between $\mu$ and $m$}\label{sec:RelationToDarboux}

The Polyakov action is defined in terms of the complex structure $I^{[\mu]}$ as
\begin{align} S_{cl} =
 &~\frac{1}{ 4\pi \alpha' } \int dx \wedge I^{[\mu]} dx,\nonumber{} \end{align}
where we make the dependence on $\mu$ and $\bar \mu$ explicit. Then it is written in terms of the Dolbeault operators as
\begin{align} S_{cl} =
 &~\frac{1}{ 4\pi \alpha' } \int ( \partial^{[\mu]} x + \bar \partial^{[\mu]} x )\wedge I^{[\mu]} ( \partial^{[\mu]} x + \bar \partial^{[\mu]} x )\nonumber{} \\ =
 &~\frac{ -i }{ 2\pi \alpha' } \int \partial^{[\mu]}x \wedge \bar \partial^{[\mu]} x.\nonumber{} \end{align}
Using eq. (\ref{dolbeault_mu}) to write the dependence on $\mu, \bar \mu$ explicitly, we get
\begin{align} S_{cl} =
 &~\frac{-i}{ 2\pi \alpha' } \int \left( \frac{ 1 + \mu \bar \mu }{ 1 - \mu \bar \mu } \partial x \bar \partial x + \frac{ dz d \bar z }{ 1 - \mu \bar \mu } \left( m \left( \frac{\partial x}{\partial z} \right)^2 + \bar m \left( \frac{ \partial x }{ \partial \bar z } \right)^2 \right) \right).\nonumber{} \end{align}
Then we compare it with the Polyakov action in terms of the $m, \bar m$ coordinates (eq. (\ref{polyakov_darboux})), which can be written in terms of the Dolbeault operators as
\begin{align} S_{cl} =
 &~\frac{ -i }{2\pi \alpha'} \left( \sqrt{ 1 + m \bar m } \partial x \bar \partial x + \frac{1}{2} dz d\bar z \left( m \left( \frac{ \partial x }{ \partial z } \right)^2 + \bar m \left( \frac{ \partial x }{ \partial \bar z } \right)^2 \right) \right).\nonumber{} \end{align}
We can then relate the coordinates $m, \bar m$ with the coordinates $\mu, \bar \mu$. The relation is
\begin{align} m \;=\;
 &\frac{ 2 \mu }{ 1 - \mu \bar \mu }\label{DarbouxViaM} \\ \mu \;=\;
 &m\over 1 + \sqrt{1 + |m|^2}\nonumber{} \end{align}
\begin{equation}
   I = {i\over 1 - |\mu|^2}
   \left(
         \begin{array}{cc}
         1 + |\mu|^2 & 2\mu \cr -2\bar{\mu} & -1 - |\mu|^2
         \end{array}
         \right)
   \end{equation}

\subsubsection{Beltrami differentials $\mu$ are holomorphic coordinates on the space of complex structures}\label{sec:BeltramiDifferentialsAreHolomorphic}

To describe a complex structure on $\Sigma$ is equivalent to saying
which functions are holomorphic. From this point of view, the definition of
Beltrami differential $\mu$ is very natural. Namely, we say that the function $f$ is holomorphic,
if:
\begin{equation}
   \left({\partial\over\partial\bar z}
         +
         \mu(z,\bar{z}) {\partial \over\partial z}\right) f(z,\bar{z}) = 0
   \end{equation}
Therefore, the definition of holomorphicity depends as a parameter on a complex-valued
$\mu(z,\bar{z})$, therefore $\mu(z,\bar{z})$ defines a holomorphic coordinate
on the infinite-dimensional space of complex structures.

Notice that  $m$ is \emph{not} a holomorphic coordinate.
We started with $m$ because it allowed a straightforward construction of
Darboux coordinates. But it is better to use $\mu$, because it is a holomorphic
coordinate.

In the rest of this Section we will explain how to replace $m,b,\bar{m},\bar{b}$ with
new Darboux coordinates $\mu,\beta,\bar{\mu},\bar{\beta}$ which agree with the complex
structure. 
It turns out that those simplifications
which we observed in coordinates $m,b,\bar{m},\bar{b}$ persist in
$\mu,\beta,\bar{\mu},\bar{\beta}$.

\subsubsection{Expansion in $\mu$}\label{sec:ExpansionInMu}

\paragraph{Expansion of $S_{\rm cl}$}\label{sec:ExpansionOfScl}

\begin{align} S_{\rm cl}\;=\;
 &\int d^2z
                \left(
                      {1 + |\mu|^2\over 1 - |\mu|^2} \partial x \bar{\partial} x
                      + {\mu\over 1 - |\mu|^2} (\partial x)^2
                      + {\bar{\mu}\over 1 - |\mu|^2} (\bar{\partial} x)^2
                      \right)\;=\;\nonumber{} \\ \;=\;
 &\int d^2z
                \left(
                      \partial x \bar{\partial} x
                      +
                      \sum\limits_{k>0}
                      2|\mu|^{2k} \partial x \bar{\partial} x
                      +
                      \sum\limits_{k\geq 0}
                      \mu^{k+1}\bar{\mu}^k (\partial x)^2
                      +
                      \sum\limits_{k\geq 0}
                      \mu^{k}\bar{\mu}^{k+1} (\bar{\partial} x)^2
                      \right)\nonumber{} \end{align}

\paragraph{Expansion of $\hat{Q}$}\label{sec:ExpansionOfQ}

It is useful to rewrite Eq. (\ref{QHat}) as follows:
\begin{align} \hat Q =
 &~\int d^2 z \bigg( \mathcal{L}_c x x^\star + \frac{1}{2} \mathcal{L}_c c c^\star + \sqrt{1 + m \bar m} ( (\bar \partial c) b +  (\partial \bar c) \bar b ) \;+ \nonumber{} \\  
 &\phantom{~\int d^2 z } 
                + \frac{1}{2} (-\partial c + \bar{\partial}\bar c) (m b - \bar{m}\bar{b})
                \label{QHatInMu} \\  
 &\phantom{~\int d^2 z } 
                                      +\frac{1}{2}
                                      \left(
                                            c (b\partial m  + \bar{b}\partial\bar{m}) +
                                            \bar{c} (b \bar{\partial} m + \bar{b}\bar{\partial}\bar{m})
                                            \right)
                                      \bigg)
                \nonumber{} \end{align}
We have:
\begin{align} b\;=\;
 &\beta {\partial \mu\over\partial m} + \bar{\beta} {\partial \bar{\mu}\over\partial m}
              =
              {1\over 2}{1 - |\mu|^2\over 1 + |\mu|^2}(\beta - \bar{\mu}^2\bar{\beta})
              \nonumber{} \\ \bar{b}\;=\;
 &\bar{\beta}{\partial\bar{\mu}\over\partial\bar{m}}
               +\beta {\partial\mu\over\partial\bar{m}}
               =
               {1\over 2}{1 - |\mu|^2\over 1 + |\mu|^2}(\bar{\beta} - \mu^2\beta)
               \nonumber{} \end{align}
In the first line of Eq. (\ref{QHatInMu}) we have:
\begin{equation}
   \sqrt{1 + |m|^2} ( (\bar \partial c) b +  (\partial \bar c) \bar b ) =
   {1\over 2}
   \left(
         (\bar{\partial}c)(\beta -\bar{\mu}^2 \bar{\beta})
         +(\partial\bar{c})(\bar{\beta} - \mu^2 \beta)
         \right)
   \end{equation}
In the middle line:
\begin{equation}
   {1\over 2}(-\partial c + \bar{\partial}\bar c)
   \left(
   \beta\left(m{\partial\over\partial m} - \bar{m}{\partial\over\partial\bar{m}}\right)\mu
   +
   \bar{\beta}\left(m{\partial\over\partial m} - \bar{m}{\partial\over\partial\bar{m}}\right)\bar{\mu}\right) =
   {1\over 2}(-\partial c + \bar{\partial}\bar c)(\beta\mu - \bar{\beta}\bar{\mu})
   \end{equation}
In the last line, we observe:
\begin{equation}
   c (b\partial m  + \bar{b}\partial\bar{m}) +
   \bar{c} (b \bar{\partial} m + \bar{b}\bar{\partial}\bar{m})
   \;=\;
   c (\beta\partial \mu  + \bar{\beta}\partial\bar{\mu}) +
   \bar{c} (\beta \bar{\partial} \mu + \bar{\beta}\bar{\partial}\bar{\mu})
   \end{equation}
This results in the following expression for $\widehat{Q}$, which is perhaps simpler
than expected:
\begin{align} \widehat{Q}\;=\;
 &\int d^2z
                \left(
                      \mathcal{L}_c x x^\star + \frac{1}{2} \mathcal{L}_c c c^\star
                      +
                      {1\over 2}
                      \left(
                            (c \stackrel{\leftrightarrow}\partial \mu) \beta +
                            \bar{\partial}(\bar c  \mu) \beta  +
                            (\bar{c} \stackrel{\leftrightarrow}{\bar\partial} \bar \mu)\bar{\beta} +
                            \partial (c\bar{\mu}) \bar{\beta}
                            \right)
                      \right)\;+\nonumber{} \\  
 &+ \int d^2 z \;
             \left(
         {1\over 2}(\bar{\partial}c)(\beta -\bar{\mu}^2 \bar{\beta})
         +{1\over 2}(\partial\bar{c})(\bar{\beta} - \mu^2 \beta)
         \right)\nonumber{} \end{align}
After imposing the condition $\bar{\partial}c = \partial\bar{c} = 0$,
the second line drops out.
Therefore, the BRST operator is linear  in \emph{both} $b,m$ and $\beta,\mu$ coordinates.
\footnote{
         This actually follows from Eq. (\ref{TransformationOfMMbar}).
         Remember that the action of $\{S_{\rm BV},\_\}$ is essentially an infinitesimal conformal
         transformations (with the parameter $c$).
         The difference between $m,\bar{m}$ and $\mu,\bar{\mu}$ is
         in a factor, which is a function of $|m|^2$.
         Eq. (\ref{TransformationOfMMbar}) says that $|m|^2$
         transforms under conformal transformations as a function. Therefore both $m$ and $\mu$
         transform as sections of  $T^{1,0}\otimes \Omega^{0,1}\;\Sigma$.
         }

\section{Derived brackets and holomorphic factorization}\label{DerivedBracketsAndContactTerms}

         Correlation functions tend to factorize as a product
         of holomorphic and antiholomorphic function of the moduli.
         In the language of Bertrami differentials this was explained in \cite{DHoker:2015qrj}.
         Here we will give a BV explanation of this phenomenon.

\subsection{Relation between contact terms and derived bracket}\label{sec:RelationBracketContact}

Let $\Psi_1$ and $\Psi_2$ be two ``gauge fermions'', \textit{i.e.} local operators
of $x,c,b$. The corresponding infinitesimal deformations of the action are:
\begin{align} S_0 \longrightarrow\;
 &S_0 + \epsilon_1\int_{\Sigma}Q_{\rm BRST} \Psi_1 + \epsilon_2\int_{\Sigma} Q_{\rm BRST} \Psi_2\nonumber{} \end{align}

\subsubsection{A relation between quadratic order and OPE of first order deformations}\label{sec:RelationToOPE}

Contact terms are delta-function terms in the OPE ${\cal O}_1(z,\bar{z}) {\cal O}_2(0,0)$:
\begin{equation}
   {\cal O}_1(z,\bar{z}) {\cal O}_2(0,0) = \delta(z,\bar{z})\Phi(0,0) + \ldots
   \end{equation}
Consider the special case when both operators are BRST-exact: ${\cal O}_j = Q_{\rm BRST} \Psi_j$.
In this case, the contact term is related to the derived bracket:
\begin{equation}
   \Phi \simeq \{\Psi_1, \Psi_2\} + Q_{\rm BRST}(\ldots)
   \label{ContactTermViaBracket}\end{equation}
Indeed:
\begin{align} \left(Q_{\rm BRST}\Psi_1(z,\bar{z})\right)\; \left(Q_{\rm BRST}\Psi_2(0,0)\right) \;=\;
 &Q_{\rm BRST}
             \left(
                   Q_{\rm BRST}\Psi_1(z,\bar{z})\;\Psi_2(0,0)
                   \right)\nonumber{} \\  
 &-\;\left(
                 Q_{\rm BRST}^2\Psi_1(z,\bar{z})
                 \right)
                \Psi_2(0,0)\nonumber{} \end{align}
The first term is BRST exact. The second term is proportional to Equations of Motion:
\begin{equation}
   Q_{\rm BRST}^2\Psi_1 = \{S_0,\Psi_1\}
   \end{equation}
where $\{\_,\_\}$ is the derived bracket.
Under the path integral, this results in the contact term $\{\Psi_1,\Psi_2\}$:
\begin{equation}
   \int [d\phi] e^{S_0} \{S_0, \Psi_1\} \Psi_2 \ldots =
   \int [d\phi] \left\{e^{S_0},\Psi_1\right\}\Psi_2 \ldots =
   \int [d\phi] e^{S_0}  \{\Psi_1,\Psi_2\}
   \label{ContactTermsDerivation}\end{equation}

\subsubsection{Contact terms and higher orders}\label{sec:ContactTermsHigherOrders}

Let us deform the Lagrangian submanifold by letting it flow along
$e^{t\{\Psi,\_\}_{\rm BV}}$ where $\Psi$ is some gauge fermion which is a functional of $\phi$
(and does not contain $\phi^{\star}$). Then:
\begin{equation}
   S_0 \mapsto S_0 + tQ_{\rm BRST}\Psi + {t^2\over 2} \{\Psi,\Psi\} + {t^3\over 6} \{\Psi,\Psi,\Psi\} + \ldots
   \label{DeformationViaBrackets}\end{equation}
Eq. (\ref{ContactTermViaBracket}) implies that $t^2\{\Psi,\Psi\}$ cancels against
the contact terms in the OPE $tQ_{\rm BRST}\Psi$. Moreover, we will now argue that
the total effect of all the contact terms is to cancel
$t^2 \{\Psi,\Psi\}$, $t^3 \{\Psi,\Psi,\Psi\}$, $\ldots$
and replace $t\Psi$ with some $\widetilde{\Psi}(t)$ in the BRST-exact term
$Q_{\rm BRST} t\Psi$. Roughly speaking:
\begin{equation}
   \exp\left(
             tQ_{\rm BRST}\Psi + {t^2\over 2} \{\Psi,\Psi\} + {t^3\over 6} \{\Psi,\Psi,\Psi\} + \ldots
             \right) =
   {\scriptstyle\tt x\atop x}
   \exp\left(
             Q_{\rm BRST}\widetilde{\Psi}(t) 
             \right)
   {\scriptstyle\tt x\atop x}
   \label{EffectiveDeformation}\end{equation}
Here ${\scriptstyle\tt x\atop x}\ldots{\scriptstyle\tt x\atop x}$ means dropping contact terms, as we now explain. 

\subsubsection{The case of bosonic string}\label{sec:ContactTermsBosonicStringCase}

Let us consider a particular case when $\Psi$ is the $b$-ghost:
\begin{equation}
   \Psi = \int m b + \int \bar{m} \bar{b}
   \end{equation}
In this case:
\begin{equation}
   S_0 \mapsto S_0 + (Q_{\rm BRST}\Psi)_{(2,0) + (0,2)} +  {1\over 2}\{\Psi,\Psi\}_{(1,1)} + {1\over 6} \{\Psi,\Psi,\Psi,\Psi\}_{(1,1)} + \ldots
   \end{equation}
where the lower index stands for the conformal dimension. All higher order terms in the deformation
have conformal dimension $(1,1)$,
and only the linear term has dimensions $(2,0)$ and $(0,2)$.
(As we explained in 
    Section \ref{sec:ConditionForSimplification},
    this actually happens in general BRST formalism under some cohomological condition.)

\subsubsection{CFT considerations}\label{sec:CFTArgument}

In this case, there is
an additional CFT-based argument (inspired by \cite{Kutasov:1988xb}),
showing that
the second order term $m\bar{m} \partial X \bar{\partial}X$ must indeed be
the contact term between $m(\partial X)^2$ and $\bar{m}(\bar{\partial}X)^2$.
The argument goes as follows. We know that the deformed
theory is conformally inviariant
(althought the action of conformal transformations changes, as we change the worldsheet
           complex structure).
Consider the exponential vertex operator:
\begin{equation}
   V_k(z,\bar{z}) = \lim_{\epsilon\rightarrow 0}
   \epsilon^{k^2}
   \exp\left(\int_{D_{\epsilon}} d^2z\; (k\cdot X(z,\bar{z}))\right)
   \label{Exponential}\end{equation}
where $D_{\epsilon}$ is the disk $|z|<\epsilon$. 

This operator remains finite when we deform the complex structure.

Let us study the effects of turning on $\mu$ on this operator.
Actually, for the conservation of momentum, we have to insert other
vertex operators $V_{p_1},\ldots, V_{p_N}$ and consider the coefficient
of $\delta(k + p_1 + \ldots p_N)$.
We can consider $\mu$ and $\bar{\mu}$ with finite support, localized
around the insertion of $V_k$, so that $\mu=\bar{\mu}=0$ at the points
of insertion of  $V_{p_1},\ldots, V_{p_N}$.
When we insert $\int \mu (\partial X)^2 \int \bar{\mu}(\bar{\partial} X)^2$,
there is logarithmic divergence due to the contact term.
It should cancel with the logarithmic divergence from
the insertion of $\int \mu\bar{\mu} (\partial X\bar{\partial} X)$.

\subsection{Holomorphic factorization}\label{sec:HolomorphicFactorization}

Consider the correlation function:
\begin{equation}
   \left\langle
   {\scriptstyle\tt x\atop x}
   e^{p_1 x(z_1,\bar{z}_1)}
   \cdots
   e^{p_m x(z_1,\bar{z}_m)}
   \exp\left(
             \int d^2z Q_{\rm BRST}(\mu b + \bar{\mu}\bar{b})
             \right)
   {\scriptstyle\tt x\atop x}
   \right\rangle
   \end{equation}
It is proportional to $\delta(p_1 + \ldots + p_m)$. Therefore we can assume that
$p_1 + \ldots + p_m = 0$.  Then:
\begin{equation}
   e^{p_1 x(z_1,\bar{z}_1)}
   \cdots
   e^{p_m x(z_1,\bar{z}_m)} =
   \exp\left(
             \sum_k \left(p_k \int_{(0,0)}^{z_k,\bar{z}_k} dz \partial x\right)
             \right)
   \exp\left(
             \sum_k \left(p_k \int_{(0,0)}^{z_k,\bar{z}_k} d\bar{z} \bar{\partial} x\right)
             \right)
   \end{equation}
Under ${\scriptstyle\tt x\atop x}\ldots{\scriptstyle\tt x\atop x}$, $\partial x$ only
talks to $T$, and $\bar{\partial} x$ only to $\bar{T}$.
Therefore we have a product of an expression depending only on $\mu$ and an expression
depending only on $\bar{\mu}$.

\subsection{The procedure of dropping contact terms}\label{sec:DroppingContactTerms}

We will now explain what we mean by ``dropping contact terms''.

Suppose that we deform the action:
$S_0 \mapsto S_0 + \int d^2 z \;\rho(z,\bar{z})\; {\cal O}(z,\bar{z})$
where $\rho$ is some function and $\cal O$ some operators. This may be thought of as
introducing $z,\bar{z}$-dependent coupling constants. Consider the correlation function
of some operators in the deformed theory:
\begin{equation}
   \left\langle
   V_1(z_1,\bar{z}_1)\cdots V_m(z_m,\bar{z}_m)
   \right\rangle_{\rho} :=\;
   \left\langle
   V_1(z_1,\bar{z}_1)\cdots V_m(z_m,\bar{z}_m)\;
   \exp\left(
             \int d^2 z \;\rho(z,\bar{z})\; {\cal O}(z,\bar{z})
             \right)
   \right\rangle_0
   \end{equation}
where $\langle \ldots \rangle_0$ is the correlation function in the undeformed theory.
Let us expand it in powers of $\rho$. Suppose that  the $N$-th power of the
expansion can be written as a sum of multiple integrals:
\begin{align}  
 &\int d^2z_1 \cdots \int d^2 z_N\; F_N(z_1,\ldots,z_N,\bar{z}_1,\ldots,\bar{z}_N)
             \;\rho(z_1,\bar{z}_1)\cdots \rho(z_N,\bar{z}_N) \;+\label{NoContactTerms} \\  
 &\int d^2z_1 \cdots \int d^2 z_{N-1}\; G_N(z_1,\ldots,z_{N-1},\bar{z}_1,\ldots,\bar{z}_{N-1})
             \;(\rho(z_1,\bar{z}_1))^2\cdots \rho(z_{N-1},\bar{z}_{N-1})\;+\label{OneDelta} \\  
 &\int d^2z_1 \cdots \int d^2 z_{N-1}\; \tilde{G}_N(z_1,\ldots,z_{N-1},\bar{z}_1,\ldots,\bar{z}_{N-1})
             \;(\rho(z_1,\bar{z}_1)\partial\rho(z_1,\bar{z}_1))\cdots \rho(z_{N-1},\bar{z}_{N-1})\;+\label{OneDeltaWithDerivative} \\  
 &\int d^2z_1 \cdots \int d^2 z_{N-2}\; H_N(z_1,\ldots,z_{N-2},\bar{z}_1,\ldots,\bar{z}_{N-2})
             \;(\rho(z_1,\bar{z}_1))^3\cdots \rho(z_{N-2},\bar{z}_{N-2})\;+\label{TwoDeltas} \\  
 &\ldots\nonumber{} \end{align}
Then we define 
\begin{align}  
 &\left\langle V_1(z_1,\bar{z}_1)\cdots V_m(z_m,\bar{z}_m)
                     \exp\left(
                               \int d^2 z \;\rho(z,\bar{z})\; {\cal O}(z,\bar{z})
                               \right)
                     \right\rangle^{\rm top} :=\;\nonumber{} \\ \;=\;
 &\Sigma_N\int d^2z_1 \cdots \int d^2 z_N \;F_N(z_1,\ldots,z_N,\bar{z}_1,\ldots,\bar{z}_N)\;
                     \rho(z_1,\bar{z}_1)\cdots \rho(z_N,\bar{z}_N)\nonumber{} \end{align}
In other words, we pick only the terms with the maximal number of integrations
over the positions of $\rho$. Equivalently, we expand the exponential
in powers of $\rho(z,\bar{z})$ and
drop the delta-functions in the OPEs ${\cal O} {\cal  O}$ and $V\cal O$:
\begin{align}  
 &\left\langle
        V_1(z_1,\bar{z}_1)\cdots V_m(z_m,\bar{z}_m)
        \exp\left(
                  \int d^2 z \;\rho(z,\bar{z})\; {\cal O}(z,\bar{z})
                  \right)
        \right\rangle^{\rm top} :=\;\nonumber{} \\ \;=\;
 &\left\langle
           {\scriptstyle\tt x\atop x}
           V_1(z_1,\bar{z}_1)\cdots V_m(z_m,\bar{z}_m)
           \exp\left(
                     \int d^2 z \;\rho(z,\bar{z})\; {\cal O}(z,\bar{z})
                     \right)
           {\scriptstyle\tt x\atop x}
           \right\rangle\nonumber{} \end{align}

\subsection{BRST invariance}\label{sec:BRSTInvarianceOfTop}

Since $Q_{\rm BRST}$ is a local symmetry of $S_0$,
this prescription is BRST-invariant:
\begin{align}  
 &\left\langle (Q_{\rm BRST}V_1)(z_1,\bar{z}_1)\cdots V_m(z_m,\bar{z}_m)
                     \exp\left(
                               \int d^2 z \;\rho(z,\bar{z})\; {\cal O}(z,\bar{z})
                               \right)
                     \right\rangle^{\rm top} \;+\nonumber{} \\  
 &\ldots \;+\nonumber{} \\  
 &\left\langle V_1(z_1,\bar{z}_1)\cdots (Q_{\rm BRST}V_m)(z_m,\bar{z}_m)
                     \exp\left(
                               \int d^2 z \;\rho(z,\bar{z})\; {\cal O}(z,\bar{z})
                               \right)
                     \right\rangle^{\rm top} \;+\nonumber{} \\  
 &\left\langle V_1(z_1,\bar{z}_1)\cdots V_m(z_m,\bar{z}_m)
                     \int d^2 z \;\rho(z,\bar{z})\;Q_{\rm BRST}{\cal O}(z,\bar{z})\;
                     \exp\left(
                               \int d^2 z \;\rho(z,\bar{z})\; {\cal O}(z,\bar{z})
                               \right)
                     \right\rangle^{\rm top} \;=\nonumber{} \\  
 &=0\nonumber{} \end{align}
Moreover, when ${\cal O} = Q_{\rm BRST} \Psi$, the deformed correlation function
\begin{equation}
   \left\langle V_1\cdots V_m \right\rangle_{\rm deformed} 
   \;:=\;
   \left\langle
   V_1 \cdots V_m 
   \exp\left(
             \int d^2 z \;\rho(z,\bar{z})\; {\cal O}(z,\bar{z})
             \right)
   \right\rangle^{\rm top}
   \end{equation}
is BRST-invariant:
\begin{equation}
   \left\langle Q_{\rm BRST}V_1\cdots V_m \right\rangle_{\rm deformed} +
   \ldots +
   \left\langle V_1\cdots Q_{\rm BRST} V_m \right\rangle_{\rm deformed} \;=\; 0
   \end{equation}
This follows from the BRST invariance of $\left\langle\ldots\right\rangle^{\rm top}$
and the fact that $Q_{\rm BRST}$ is nilpotent on-shell. They key point is that
the terms proportional to the equations of motion make multiple integrals collapse to the
lower multiple integrals, and therefore do not contribute to the $\left\langle\cdots\right\rangle^{\rm top}$.

\subsection{Relation between prescriptions}\label{sec:RelationBetweenPrescriptions}

Exists operator ${\cal O}_{\rho}$:
\begin{equation}
   {\cal O}_{\rho}(z,\bar{z}) = \rho(z,\bar{z}){\cal O}(z,\bar{z}) + O(\rho^2)
   \end{equation}
and a map ${\cal W}_{\rho}$ on the space of operators:
\begin{align}  
 &V\mapsto {\cal W}_{\rho}[V]\nonumber{} \\  
 &{\cal W}_{\rho}[V] = V + O(\rho)\nonumber{} \end{align}
such that:
\begin{align}  
 &\left\langle
       V_1 \cdots V_m 
       \exp\left(
                 \int d^2 z \;\rho(z,\bar{z})\; {\cal O}(z,\bar{z})
                 \right)
       \right\rangle^{\rm top}=\nonumber{} \\ \;=\;
 &\left\langle
           {\cal W}_{\rho}[V_1](z_1,\bar{z}_1) \cdots {\cal W}_{\rho}[V_m](z_m,\bar{z}_m)
           \exp\left(
                     \int d^2 z \; {\cal O}_{\rho}(z,\bar{z})
                     \right)
           \right\rangle\label{DeformedCorrelatorOfW} \end{align}
Both ${\cal O}_{\rho}(z,\bar{z})$ and ${\cal W}_{\rho}[V](z,\bar{z})$ depend on
$\rho(z,\bar{z})$ and its derivatives with respect to $z$, $\bar{z}$.
The BRST operator induces a map $Q_{\rho}$ on the space of operators
such that:
\begin{align}  
 &Q_{\rho} {\cal W}_{\rho}[V] = {\cal W}_{\rho} [Q_{\rm BRST} V]\nonumber{} \\  
 &Q_{\rho} {\cal W}_{\rho}[V] = Q_{\rm BRST} V + O(\rho)\nonumber{} \end{align}
If ${\cal O} = Q_{\rm BRST} \Psi$, then by construction $Q_{\rho}$
is a symmetry of the correlation functions, and therefore it corresponds to a conserved charge.
The $Q_{\rho}$ is the BRST transformation
(defined in Section \ref{sec:BRST})
of the theory deformed by the insertion of 
$\exp\left(
             \int d^2 z \; {\cal O}_{\rho}(z,\bar{z})
             \right)$
--- the second line of Eq. (\ref{DeformedCorrelatorOfW}).

\subsection{Explicit computation using Wick theorem}\label{sec:WickTheorem}

We will now do an explicit computation for the matter part of the theory.

\subsubsection{Action and propagator}\label{sec:ActionAndPropagator}

The undeformed path integral is:

\begin{equation}
   \int [dx] \exp\left( - {1\over \pi\alpha'} \int d^2z \,(\bar{\partial}x \partial x)\right)
   \;=\;
   \int [dx] \exp\left(
                       - {1\over 2}
                       {1\over \pi\alpha'}
                       \int d^2z \;
                       (\partial x, \bar{\partial}x)
                       \left(
                             \begin{array}{cc}
                             0 & 1 \cr 1 & 0
                             \end{array}
                             \right)
                       \left(
                             \begin{array}{c}
                             \partial x \cr \bar{\partial} x
                             \end{array}
                             \right)
                       \right)
   \end{equation}
With this action:
\begin{equation}
   \left\langle \bar{\partial}x(z,\bar{z}) \partial x (0,0) \right\rangle
   = {\pi \alpha'\over 2} \delta^2(z,\bar{z})
   \end{equation}
with notations: $z = x + iy$, $\delta^2(z,\bar{z}) = \delta(x)\delta(y)$.
\begin{equation}
   \left\langle
   \left(\begin{array}{cc} \partial x(z,\bar{z}) \cr \bar{\partial} x(z,\bar{z}) \end{array}\right)
   \left(\partial x(0,0)\,,\,\bar{\partial} x(0,0)\right)
   \right\rangle_{\tt contact\; terms} =  {\pi \alpha'\over 2}
   \left(\begin{array}{cc} 0 & \delta^2(z,\bar{z}) \cr  \delta^2(z,\bar{z}) & 0 \end{array}\right)
   \end{equation}
Complex structure deformation corresponds to the insertion of:
\begin{equation}
   \exp
   \left(
         -{1\over 2}
         {1\over \pi\alpha'}
         \int d^2z \;
         (\partial x, \bar{\partial}x)
         \left(
               \begin{array}{cc}
               m & -1 + \sqrt{1 + |m|^2} \cr
               -1 + \sqrt{1+|m|^2} & \bar{m}
               \end{array}
               \right)
         \left(
               \begin{array}{c}
               \partial x \cr \bar{\partial} x
               \end{array}
               \right)
         \right)
   \end{equation}

\subsubsection{Normal ordering of exponentials of quadratic expressions}\label{sec:NormalOrdering}

Consider a linear space with coordinates $x^1,\ldots,x^N$ and a symmetric matrices $A$ and $G$.
Notice that:
\begin{equation}
   \exp\left({1\over 2}G^{ij}{\partial\over\partial x^i} {\partial\over\partial x^j}\right)
   \exp\left(- {1\over 2}A_{ij}x^i x^j\right)
   =
   \exp\left(- {1\over 2}[A({\bf 1} + GA)^{-1}]_{ij}x^i x^j\right)
   \label{ContractionsInExpOfQuadratic}\end{equation}
\begin{align} A \;=\;
 &{1\over \pi\alpha'}
        \left(
              \begin{array}{cc}
              m & -1 + \sqrt{1 + |m|^2} \cr
              -1 + \sqrt{1 + |m|^2} & \bar{m}
              \end{array}
              \right)\nonumber{} \\ G\;=\;
 &{\pi\alpha'\over 2}\left(\begin{array}{cc} 0 & 1 \cr 1 & 0 \end{array}\right)\nonumber{} \end{align}
\begin{align}  
 &\left(
              1
              +
              {1\over 2}\left(
                              \begin{array}{cc}
                              -1 + \sqrt{1 + |m|^2} & \bar{m} \cr
                              m & -1 + \sqrt{1 + |m|^2}
                              \end{array}
                              \right)
              \right)^{-1}\;=\;\nonumber{} \\ \;=\;
 &\left(
                 \begin{array}{cc}
                 2 & - {2\bar{m}\over 1 + \sqrt{1+|m|^2}} \cr
                 - {2m\over 1 + \sqrt{1+|m|^2}} & 2
                 \end{array}
                 \right)\nonumber{} \end{align}
\begin{align}  
 &A\left(
              1
              +
              {1\over 2}\left(
                              \begin{array}{cc}
                              -1 + \sqrt{1 + |m|^2} & \bar{m} \cr
                              m & -1 + \sqrt{1 + |m|^2}
                              \end{array}
                              \right)
              \right)^{-1}\;=\;\nonumber{} \\ \;=\;
 &{2\over \pi\alpha'}
        \left(
              \begin{array}{cc}
              {m\over 1 + \sqrt{1 + |m|^2}} & 0 \cr
              0 & {\bar{m}\over 1 + \sqrt{1 + |m|^2}}
              \end{array}
              \right)\nonumber{} \\ \;=\;
 &{2\over \pi\alpha'}
           \left(
                 \begin{array}{cc}
                 {- 1 + \sqrt{1 + |m|^2} \over \bar{m}} & 0 \cr
                 0 & {- 1 + \sqrt{1 + |m|^2}\over m}
                 \end{array}
                 \right)\label{GeometricalProgressionResummed} \end{align}
Therefore:
\begin{align}  
 &\exp
        \left[
              -{1\over \pi\alpha'}
              \int d^2z
              \left(
                    (-1 + \sqrt{1 + |m|^2})(\bar{\partial} x\partial x)
                    + {1\over 2} m(\partial x)^2 + {1\over 2}\bar{m}(\bar{\partial}x)^2
                    \right)
              \right]
        \;=\;\nonumber{} \\ \;=\;
 &{\scriptstyle\tt x\atop \scriptstyle\tt x}
           \exp
           \left[
                 -{1\over \pi\alpha'}
                 \int d^2z\left(
                                {m\over 1 + \sqrt{1 + |m|^2}}(\partial x)^2
                                +
                                {\bar{m}\over 1 + \sqrt{1 + |m|^2}}(\bar{\partial} x)^2
                                \right)
                 \right]{\scriptstyle\tt x\atop \scriptstyle\tt x}\label{ContactWick} \\ \;=\;
 &{\scriptstyle\tt x\atop \scriptstyle\tt x}
           \exp
           \left[
                 -{1\over \pi\alpha'}
                 \int d^2z\left(
                                \mu(\partial x)^2
                                +
                                \bar{\mu}(\bar{\partial} x)^2
                                \right)
                 \right]{\scriptstyle\tt x\atop \scriptstyle\tt x}\label{ContactWickViaMu} \end{align}
where ${\scriptstyle\tt x\atop x}\ldots{\scriptstyle\tt x\atop x}$ means dropping contact terms, and $\mu$ is defined in Eq. (\ref{SplitOfT}).
The appearance of  $\mu$ in Eq. (\ref{ContactWickViaMu}) can be understood in the following way.
In the deformed theory, we know that there should be no contact terms in the OPE of
$\partial x + \bar{\mu}\bar{\partial x}$ with $\partial x + \bar{\mu}\bar{\partial} x$.
(This is because $\partial^{[\mu]}x = f(z,\bar{z})(\partial + \bar{\mu}\bar{\partial})x$.)
The contractions between $\bar{\mu}\bar{\partial x}$ and $\partial x$ cancel against the contraction with ${1\over 2}\bar{\mu}(\bar{\partial} x)^2$ brought down
from the action. 
\footnote{
Essentially, we used:
\begin{equation}
   \mbox{det}\;
   \left(
         \begin{array}{cc} \sqrt{1 + |m|^2} & \bar{m} \cr m & \sqrt{1 + |m|^2} \end{array}
         \right)
   =1
   \end{equation}
For any $2\times 2$ matrix $U$ such that $\mbox{det} \,U =1$:
\begin{equation}
   \mbox{tr}{{\bf 1} - U\over {\bf 1} + U} = 0
   \end{equation}
implying the vanishing of the off-diagonal terms in Eq. (\ref{GeometricalProgressionResummed}).
Notice that the transformation $U\mapsto {{\bf 1} - U\over {\bf 1} + U}$ squares to identity.
}

Eq (\ref{ContactWick}) shows that not only at the second order in $m$,
but to all orders, the term $(\bar{\partial}x \partial x)$
can be interpreted as the effect of contact terms.
Notice that Eq. (\ref{ContractionsInExpOfQuadratic}) can be ``inverted'':
\begin{equation}
   \exp\left(- {1\over 2}A_{ij}x^i x^j\right)
   =
   \exp\left(- {1\over 2}G^{ij}{\partial\over\partial x^i} {\partial\over\partial x^j}\right)
   \exp\left(- {1\over 2}[A({\bf 1} + GA)^{-1}]_{ij}x^i x^j\right)
   \label{ContractionsInExpOfQuadraticInverted}\end{equation}

\subsection{Summary}\label{sec:HolomorphicFactorizationSummary}

The worldsheet metric depends on the complex structure, which can be parametrized
by the Beltrami differentials $\mu$ and $\bar{\mu}$.
Here we have shown that the dependence of correlation functions on $\mu$ and $\bar{\mu}$
can be effectively computed by inserting the exponential of a linear function of
$\mu$ and $\bar{\mu}$:
\begin{equation}
   \exp(\int \mu T + \bar{\mu} \overline{T})
   \end{equation}
and dropping some contact terms. We explained this simplification from the point
of view of BV/BRST formalism.

\section*{Acknowledgments}

This work was supported in part by FAPESP thematic grants 2016/01343-7 and 2019/21281-4.
A.M. wants to thank Andrey Losev for useful discussions.

\def\cprime{$'$} \def\cprime{$'$}
\providecommand{\href}[2]{#2}\begingroup\raggedright\endgroup
\end{document}